# An exploration of advanced X-Divertor scenarios on ITER

B Covele, P Valanju, M Kotschenreuther, and S Mahajan
University of Texas at Austin Institute for Fusion Studies

**Abstract.** It is found that the X-Divertor (XD) configuration [1-3] can be made with the conventional PF coil set on ITER[4], where all PF coils are outside the TF coils. Desirable configurations are possible where the PF currents are below the present maximum design limits on ITER, and where the baseline divertor cassette is used. It is possible that the XD could be used to assist in high-power operation on ITER, but some further issues need examination. Note that the increased major radius of the Super X-Divertor (SXD) [5-8] is not a feature of the XD geometry. In addition, we present an XD configuration for K-DEMO [9], to demonstrate that it is also possible to attain the XD configuration in advanced tokamak reactors with all PF coils outside the TF coils. The results given here for the XD are far more encouraging than recent calculations by Lackner and Zohm [10] for the Snowflake [11,12], where the required high PF currents represent a major technological challenge. The magnetic field structure in the outboard divertor SOL [13] in the recently created XD configurations reproduces what was presented in the earlier XD papers [1-3]. Consequently, the same advantages accrue, but no close-in PF coils are employed.

## 1. Introduction

Power exhaust on fusion reactors beyond ITER is extremely challenging [1-3]. Recent projections of SOL widths [14,15] accentuate the difficulties, and may also pose difficulties for ITER's operation. The X-Divertor (XD) was proposed for both ITER and reactors to improve power exhaust by increasing flux expansion and line length [1-3], and to allow a higher level of acceptable detachment [1,13]. In the initial publications for the XD, the divertor magnetic geometry was created with the assistance of special PF coils relatively closer to the plasma. Here, we show that XDs can be created without such special coils, and when all PF coils are located outside superconducting TF coils, with moderate PF coil currents. The magnetic geometry in the physically relevant outboard SOL region is the same as earlier XD cases, and so the same advantages accrue [13]. See Fig. 1 for ITER cases and Fig. 2 for AT reactor cases. Our results for the XD are far more encouraging than related work of Lackner and Zohm, [10], who considered implementing a later divertor magnetic divertor geometry, the Snowflake [11,12], on ITER, but found that the required high PF currents were well above the ITER design values and were deemed a major technological challenge. Fortunately, we find that an XD may be created with considerably lower PF currents than a Snowflake. In the standard PF coil set of ITER, very interesting XDs can be created with coil currents below the nominal maximum design values of ITER, and even using the baseline divertor cassette. Our results strongly motivate more complete future analysis of the feasibility and benefits of the XD for ITER and future reactors, including issues beyond the creation of the equilibrium. Divertor operation with an XD on ITER could both assist in the high-power ITER phase, and also allow a test of the XD prior to implementation on a future demonstration reactor (DEMO). As a possible example of the latter, we present an XD for K-DEMO [9] with Advanced Tokamak (AT) operation, where all PF coils are outside the TF coils. These examples indicate the magnetic feasibility of XDs for many superconducting tokamaks with standard PF sets. To avoid confusion, we explicitly note that in this paper we will only discuss the XD, and not the Super X-Divertor (SXD) which came several years later. In the SXD, the major radius of the XD was increased [5-8] to further improve performance, but this is not a feature of the XD.

## 2. Methods

The free boundary MHD equilibrium code CORSICA [16] is used for all cases. For definiteness, we assume an outboard SOL width of 2 mm. For ITER, we consider only the baseline Ohmic phase here. The core equilibria have $<\beta_N> = 1.8$, $l_i(3) = 0.83$ and $p(0)/<p> = 1.9$. We use the PF and CS coil locations from Ref. 17, and the first wall location from Ref. 18. No coil current is allowed to exceed its maximum design limit [17]. Divertor configurations with an XD only on the outboard side are considered – this is where the heat flux is most serious. For K-DEMO, we consider both single-null and double-null cases. The single-null cases have an XD on both the inboard and the outboard divertor legs. We consider core plasma equilibria characteristics of AT operation, with hollow current profiles ($l_i(1) = 0.7$, $l_i(3) = 0.5$), and $<\beta_N> = 3.9$ [19].

Briefly, an X-Divertor is created when a second x-point is introduced downstream in the divertor SOL, such that the magnetic field lines flare more than a standard divertor at the target plate. A quantitative measure of this flaring is given by the Divertor Index (DI); an X-Divertor, by definition, has a DI greater than 1 [13].

## 3. ITER results

The case of ITER is made more challenging because the PF coils and wall location are already fixed, and XD operation was not envisaged. For the standard divertor, the closest distance from the separatrix to the wall is only 15 cm, or 7% of the minor radius. With the ITER PF coil set, the shape of the core plasma with an XD is slightly altered. It is necessary to slightly reduce the minor radius to maintain the same minimum distance of the core plasma from the wall. The plasma confinement time estimated from ITER98H(y,2) [20] is not materially changed, as seen in Table 1.

We presume that initial operation will be with the baseline divertor cassette for a standard divertor configuration (SD). To utilize the same divertor target apparatus, the strike points for the XD must be at or very near those of the SD. Also, the SOL must safely clear the dome structure designed for the SD, so we keep the position of the core x-point the same as the SD case.

In Fig. 1, we show several ITER X-Divertor cases and compare them to the 2004 example XD as well as the standard divertor. These example scenarios represent operating points in a continuous parameter space from SD to XD, bounded only by the practical constraints of ITER. All ITER PF coil currents are below the design maximum values. Numerical parameters are given in Table 1. The continuous nature of the XD parameter space allows for an incremental application of the XD to ITER, which could be advantageous if the standard divertor geometry and the baseline cassette engender outboard heat fluxes which are only modestly higher than desired. This freedom is available with the existing PF coils by only changing the PF currents, and a commensurate, slight changing of the core plasma shape.

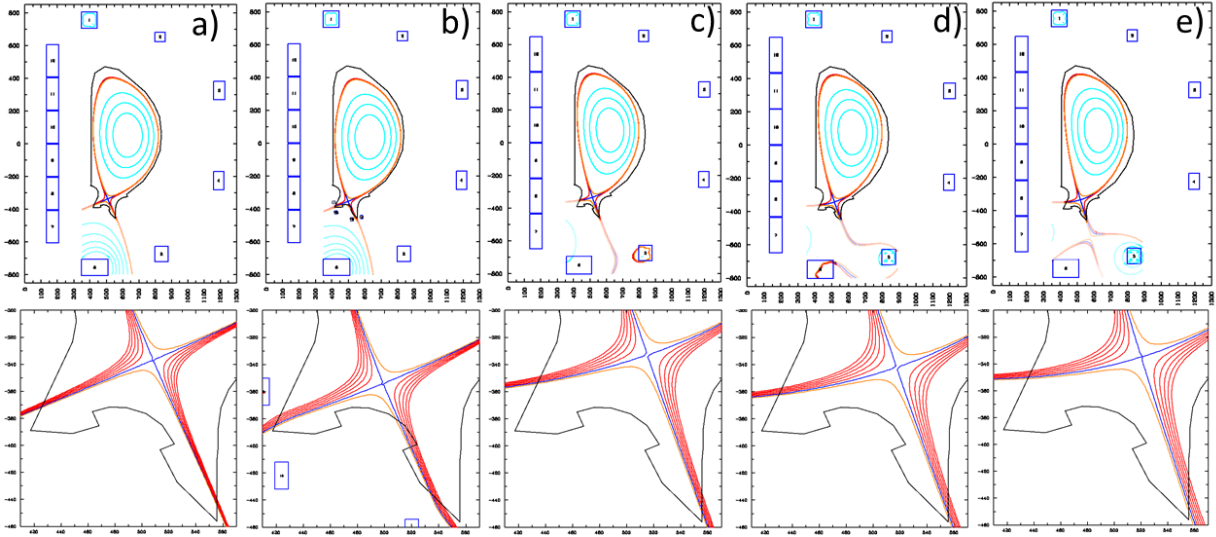

**Fig. 1.** ITER X-Divertors. (a) The ITER standard divertor baseline equilibrium. (b) The original 2004 version of the ITER XD with special PF coils near the targets. The divertor does not clear the dome structure. (c) A 2013 XD with outer flux expansion optimized for an incident B field at 1 degree at the outer target. (d) A 2013 XD with maximal outer flux expansion within coil current limits and with 15 MA of plasma current. (e) A 2013 XD with maximal outer flux expansion within coil current limits and with 14 MA of plasma current.

**Table 1.** Plasma and divertor parameters for the X-Divertor equilibria in Fig. 1. The minor radius was reduced in the 2013 XD plasmas to maintain 15 cm of clearance from the first wall. Confinement times are computed using the ITER98(y,2) scaling, with a density 80% of the Greenwald limit [20] and an assumed heating power of 120MW.

| ITER X-Divertor Scenarios | Standard Divertor | 2004 Dual XD | 1-degree-limited XD | Max. expanded XD (15 MA) | Max. expanded XD (14 MA) |
|---|---|---|---|---|---|
| Plasma current (MA) | 15.0 | 15.0 | 15.0 | 15.0 | 14.0 |
| Minor radius (cm) | 200 | 200 | 186 | 186 | 183 |
| Elongation | 1.84 | 1.86 | 2.02 | 2.05 | 2.06 |
| Upper triangularity | 0.44 | 0.37 | 0.48 | 0.49 | 0.54 |
| Lower triangularity | 0.52 | 0.59 | 0.47 | 0.48 | 0.48 |
| Outer flux exp. | 2.4 | 5.2 | 5.6 | 7.1 | 9.3 |
| Outer DI | 1.05 | 1.88 | 1.64 | 1.74 | 2.04 |
| Outer conn. length (m) | 86.1 | 97.8 | 118.5 | 128.1 | 147.0 |
| Confinement time (s) | 3.00 | 3.03 | 3.24 | 3.28 | 3.03 |
| Min. wall clearance (cm) | 12.5 | 8.0 | 15.1 | 15.0 | 15.0 |
| Avg. wall clearance (cm) | 26.5 | 26.2 | 32.6 | 31.6 | 33.9 |

\* 2004 XD results (w/ special PF coils)   \* 2013 XD results (ITER coils)

Notice that in this paper, we have concentrated on "optimizing" the heat handling capacity of the outer divertor. For all cases presented, the inner divertor does show an increased flux expansion of ~ 30-50% over the SD. However we are still working on configurations that optimize, simultaneously, the outer and inner divertors.

In Fig. 1c, we display an ITER case where the angle of the total magnetic field with the plate is 1 degree, and for which the outer flux expansion is increased by a factor ~2.3 compared to the standard ITER divertor. It seems, however, that the 1-degree requirement may seriously reduce the available wetted area for the current ITER design (the monoblocks are oriented to hide edges). In such a situation, one has to live with a reduced flux expansion, and improved divertor action will require a higher degree of detachment in the peak heat flux regions. Interestingly enough, the possibility of operating at higher levels of detachment (without degrading H-mode confinement) is one of the key features of XDs with higher values of Divertor Index (DI) [13]. Since the divertor cassette is designed to be replaceable, it may be very advantageous to redesign the cassette, modifying the orientation of the monoblock surfaces to the plasma, so that the enhanced flux expansion could be exploited to boost divertor action.

A robust trend that we have observed is that the PF coil currents increase as the second x-point is moved closer to the core x-point. This is in qualitative agreement with the results of Lackner and Zohm, who found that PF coil currents were very far beyond the ITER design limits when the second x-point coalesces with the core x-point, which is a pure Snowflake. Space does not permit a more quantitative discussion of coil current vs. x-point distance; this will be examined in future publications. For now, we note that, within the ITER design limits for PF currents, higher DI is possible only for plasmas with reduced plasma current. A case of this type is also given in Fig. 1e.

Important issues remain to be examined in the future, including (1) higher elongation and higher average distance from the wall increase the vertical instability growth rate and may challenge the control system; (2) off-normal events such as disruptions and coil failures need analysis; and (3) the flux swing (volt-seconds) possible with XD cases needs analysis.

## 4. K-DEMO results

Since the K-DEMO design is not yet fixed, there is significant freedom in the placement of the PF coils. We have placed the coils consistent with a vertical maintenance scheme [21]. We consider two cases: (1) a single-null with XDs on both outboard and inboard legs, and (2) a symmetric double-null with outboard XDs only. With a symmetric double-null, about an order of magnitude less power flows to the inboard, so an XD may not be needed on those legs. As can be seen in Fig. 2, XD geometries are possible with all PF coils outside the TF coils.

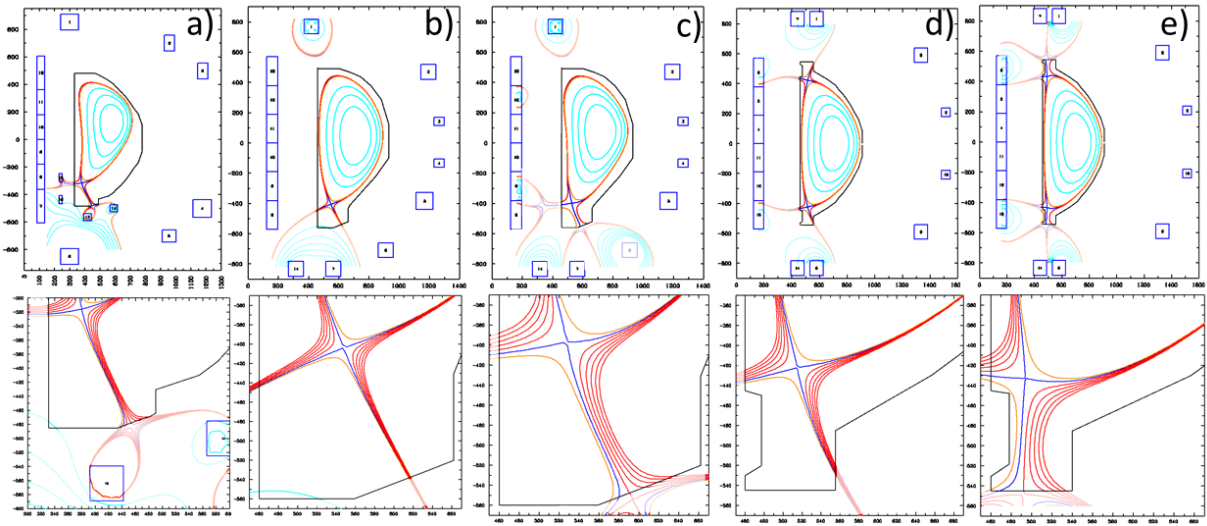

**Figure 2.** K-DEMO X-Divertors, compared to the 2004 XD on CREST. (a) 2004 CREST dual XD (XD on inboard and outboard legs) with special PF coils. (b) A standard divertor baseline equilibrium for a K-DEMO single-null plasma. (c) A conceptual design for a K-DEMO single-null dual XD with accommodation for vertical maintenance. (d) A standard divertor baseline equilibrium for a K-DEMO double-null plasma. (e) A K-DEMO double-null with an XD on the outboard leg only. Coil sizes and locations closely resemble current K-DEMO design specifications.

The K-DEMO XD in Fig. 2c (single-null plasma), with no special PF coils, has essentially the same geometry as the 2004 CREST [22] XD in Fig. 2a, which used special PF coils near the targets. Both equilibria have XDs on both the inboard and outboard legs. Furthermore, a K-DEMO XD for a double-null plasma (Fig. 2e) was designed to closely conform with current K-DEMO PF coil specifications.

## 5. Conclusions

Using ITER and K-DEMO as examples, we find that X-Divertors are possible for superconducting tokamaks with PF coils located outside the TF coils. Hence, further research on XDs is warranted: experimental tests on existing devices, investigations of detachment, and design work for burning plasmas. Such operation could potentially significantly assist in ITER's high-power phase, and also be an invaluable demonstration of the XD prior to DEMO.